\documentstyle[prb,aps, multicol,epsf]{revtex}

\begin{document}
\draft

\title{Thermodynamics of the first--order vortex lattice\\ melting
  transition in YBa$_2$Cu$_3$O$_{7-\delta}$}

\author{Matthew J.\ W.\ Dodgson,$^{a\,}$ Vadim B.\ 
  Geshkenbein,$^{a\,,b}$  Henrik Nordborg,$^{a\,,c}$ and Gianni
  Blatter$^{a\,}$ }

\address{$^{a\,}$Theoretische Physik, ETH-H\"onggerberg, CH-8093
  Z\"urich, Switzerland}

\address{$^{b\,}$L. D. Landau Institute for Theoretical Physics,
  117940 Moscow, Russia}

\address{$^{c\,}$Argonne National Laboratory, 
9700 South Cass Avenue, Argonne, IL 60439, USA}

\date{December 12, 1997}
\maketitle
\begin{abstract}

Using the London approximation within the high field
scaling regime, we
calculate the jump in the specific heat $\Delta c$ at the first--order melting 
transition of
the vortex lattice in YBa$_2$Cu$_3$O$_{7-\delta}$. This has recently been
measured 
[A.~Schilling {\it et al.}, Phys.\ Rev.\ Lett.\ {\bf 78}, 4833  (1997)] 
and reported to be at least 100 times
higher than expected from the fluctuations of field induced vortices
alone. We demonstrate how the
 correct treatment of the temperature dependence of the model
parameters, which are singular at the mean--field $B_{c2}$ line, leads to
good agreement between the predictions of the London model and
the size of the experimental jump. In addition, we consider the
changes in the slopes of the magnetization $\Delta(\partial M/\partial T)$
and $\Delta(\partial M/\partial H)$ at the transition.
Using continuum anisotropic scaling theory we demonstrate the consistency of
measurements at different angles of the magnetic field with respect to the
crystal {\bf c}-axis.

\end{abstract}


\begin{multicols}{2}
\narrowtext

\section{Introduction}\label{sec:intro}
In a recent paper\cite{us} we 
calculated the size of the jumps in the entropy and
magnetization at the vortex--lattice melting transition within the London
model. We have shown that, by correctly treating the temperature dependence of
the model parameters, and knowing the volume of the relevant fluctuation
degrees of freedom, good agreement is obtained with the experimental
results in both YBa$_2$Cu$_3$O$_{7-\delta}$ (YBCO) and
Bi$_2$Sr$_2$Ca$_1$Cu$_2$O$_8$ (BiSCCO) superconductors. 
The analysis contains only one unknown
number, which
can be taken from numerical simulations. The importance of
these results is that they settle  a controversy of recent years: it had been
thought 
that the observed jumps of order 1 $k_B$ per vortex per 
layer\cite{Zeldov,Schilling} are 
incompatible with a simple melting scenario based on fluctuations of field
induced vortices alone. 
Our analysis has demonstrated that there is no incompatibility because the
temperature dependence of the model parameters reflects the underlying 
microscopic degrees of freedom. This allows 
the correct size of jumps to be found
without having to explicitly include extra fluctuations.

Another characteristic of the melting transition is the jump in the specific
heat capacity $\Delta c=T\Delta(\partial s/\partial T)=-T\Delta
(\partial^2g/\partial T^2)$, where $g$ is the Gibbs free energy density.
There have been
careful measurements of this in YBCO,
\cite{Schilling2,Junod} and in Ref.~\onlinecite{Schilling2} 
it was claimed that
the step in the  specific
heat is at least one hundred times too large to be explained by the extra
fluctuations of the
translational degrees of freedom in the vortex liquid. The main purpose of
this paper is to explain the size of the specific heat step using the London
model, following the same approach of Ref.~\onlinecite{us}.
We emphasize that at the relevant melting fields in the
YBCO system, the penetration depth $\lambda(T)$ is much larger than the
distance between vortices $a_0\sim (\Phi_0/B)^{1/2}$ ($B$ is the magnetic
induction, or flux density, and $\Phi_0$ is the flux quantum). 
In this regime the London model has simple scaling properties, which may be
used to find the exact form of the jumps at the transition.

In the next section we calculate the heat capacity within the London model,
and compare our estimated jump at the transition
with the experimental values. In Sec.~\ref{sec:mag}
we consider the jumps in the magnetization slopes 
\hbox{$\Delta(\partial M/\partial T)=-\Delta(\partial^2 g/
\partial H\partial T)$}
and \hbox{$\Delta(\partial M/\partial H)=-\Delta(\partial^2 g/\partial H^2)$}
and find results consistent
with experimental values given by Welp {\em et al}.\cite{Welp} 
We also take the opportunity to compare the jumps in the entropy and the
specific heat between a system at constant external field (the experimental
scenario) and a system at constant vortex density (the case for most
simulations). 
Finally in Sec.~\ref{sec:angle}
we consider the effects of rotating the magnetic field
away from the {\bf c}-axis, as has been done in recent specific heat
measurements.\cite{Schilling2,Schilling3}

\section{The heat capacity in the London model}\label{sec:heat}

We first consider an isotropic superconductor in the mixed state. Within
the London approximation, the magnitude of the superconducting order parameter
is taken to be constant, except within the vortex cores, which are assumed to
be much smaller than the distance between vortices. These assumptions will hold
at low enough fields below the upper critical field $H_{c2}$. The free energy
of the system can then be expressed as a sum over pairwise interactions
between vortex segments,\cite{review} i.e.
\begin{equation}
  \label{eq:London}
  {\cal F}_L[\{ \hbox{\bf r}_\mu \}]=\frac{\varepsilon_0}{2} \sum_{\mu\nu}\int 
d\hbox{\bf r}_\mu \cdot d\hbox{\bf r}_\nu
\frac{ e^{-| \hbox{\bf r}_\mu-\hbox{\bf r}_\nu  |/\lambda}}
{|\hbox{\bf r}_\mu-\hbox{\bf r}_\nu |}.
\end{equation}
Disregarding the small distance cut-off, $\xi$, 
 there are two length scales in this problem: the London
screening length $\lambda$ and the average distance between vortices $a_0$.
The energy scale per unit length $\varepsilon_0$ is equal to 
$(\Phi_0/4\pi\lambda)^2$.
It is simple to include a uniaxial
anisotropy into the model when the effective mass in the field
direction is different from the other two perpendicular directions: For a
ratio of $m_{ab}/m_c=\varepsilon^2$, 
the lengths in the field direction are
scaled by $\varepsilon$.

The experiments of interest are carried out at constant external field
$H$. For the case of large field melting, as in YBCO, the
magnetization is small and $B\approx H$.
In this regime we have $a_0\ll\lambda$ and the London free
energy takes a simple scaling form,\cite{review}
\begin{equation}
  \label{eq:scaling}
  {\cal F}_L[\{ \hbox{\bf r}_\mu \}]\approx \varepsilon a_0\varepsilon_0 
f_L[\{ \hbox{\bf s}_\mu \}],
\end{equation}
where $\hbox{\bf s}_\mu=(x_\mu,y_\mu,z_\mu/\varepsilon)/a_0$ 
are dimensionless position vectors and
$f_L[\{ \hbox{\bf s}_\mu \}]$ is a dimensionless functional independent of
$\lambda$. The thermodynamic properties of the system are determined by the
partition function, $Z= \hbox{Tr} 
\exp{\left( -\varepsilon a_0\varepsilon_0 
f_L[\{ \hbox{\bf s}_\mu \}]/T\right)}$. Notice that
one may think of $\tau=T/\varepsilon a_0\varepsilon_0$ 
as an effective dimensionless
temperature. From the standard thermodynamic relations,
$F=-T\ln Z$, 
$S=-(\partial F/\partial T)_B$, and
$E=F+TS$, the entropy is
\begin{equation}
  \label{eq:entropy}
  S=-\frac{F}{T} +\frac{\langle{\cal F}\rangle}{T} -
\left\langle\frac{\partial{\cal F}}{\partial T}\right\rangle.
\end{equation}
The contribution $S_0=(\langle{\cal F}\rangle -F)/T$ may be thought of as the
configurational entropy of the coarse grained model, while the last term in
(\ref{eq:entropy}) represents additional contributions from the underlying
microscopic degrees of freedom that appear in the temperature dependence of
the model parameters.

In the London model we must include
 the
temperature dependence of the energy scale $\varepsilon_0(T)$ which we show in
Ref.~\onlinecite{us} to be important for an adequate determination
of the jump in the entropy 
 at the melting transition. 
Using the scaling of Eq.~(\ref{eq:scaling}), the entropy jump is\cite{us}
\begin{equation}
  \label{eq:entjump}
  \Delta S=\left(
  1-\frac{T}{\varepsilon_0}\frac{d\varepsilon_0}{dT} \right)\Delta S_0
=\frac{(1+t^2)}{(1-t^2)}\Delta S_0.
\end{equation}
The last form in Eq.~(\ref{eq:entjump}) uses the phenomenological
temperature dependence of the penetration depth,
$\lambda(T)^2=\lambda_0^2/(1-t^2)$, where $t=T/T_c$ is the reduced
temperature. The length $\lambda_0$ is found from 
fitting this
form to the slope of $1/\lambda(T)^2$ as the zero field
transition temperature $T_c$ is approached,\cite{Hardy} which for YBCO gives 
$\lambda_0=$1300~\AA.
\begin{figure}[b]
\centerline{\epsfxsize= 8.5cm\epsfbox{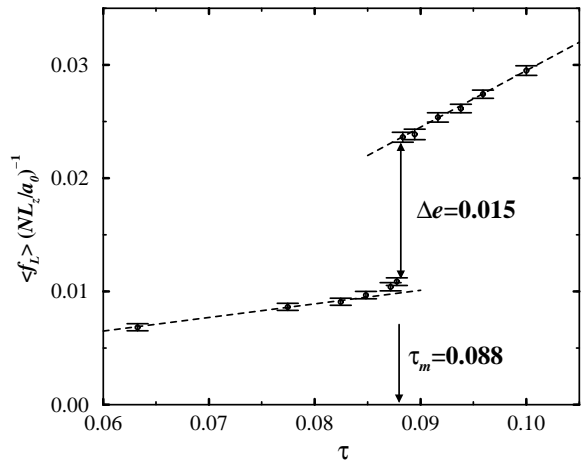}}
\caption{ The configurational average of the London free energy for
  temperatures near melting, as measured
  in the simulations of Ref.~10
 for a system of 81 lines, in
the incompressible limit $a_0\ll\lambda$. 
The dashed lines are straight line fits for
the solid and liquid phases.}
\label{fig:1}
\end{figure}
 
The configurational entropy jump $\Delta S_0$
 may be determined from numerical simulations.
In the recent simulations of Nordborg and Blatter\cite{NordborgBlatter} 
the expectation value of 
$f_L[\{ \hbox{\bf s}_\mu \}]$ is 
measured at different temperatures for a system at infinite $\lambda$, the
results of which we show in Fig.~\ref{fig:1}. 
This simulation uses the approximation of
only including interactions within the same plane perpendicular to the
external field, i.e., ``retarded'' interactions (in the Bose
language\cite{review}) 
are ignored. The volume of the system is $V= L_zNa_0^2\sqrt{3}/2$ with the 
number of vortices given by $N$, and the system length in the field direction
equal to $L_z$ (periodic boundary conditions were used in all three
directions).
Inspection of Fig.~\ref{fig:1} gives
\begin{equation}
  \label{eq:delfl}
\Delta S_0=
  \frac{\Delta \langle f_L\rangle}{\tau_m}\approx 0.17 NL_z/a_0,
\end{equation}
where the transition occurs at \mbox{$\tau=\tau_m=0.088$}.
Eq.~(\ref{eq:delfl}) shows that the jump in configurational
entropy is a constant per elementary
degree of freedom: $\Delta S_0=0.17\, V/V_{\rm edf}$, where 
$V_{\rm edf} =\varepsilon a_0\Phi_0/B\approx \varepsilon a_0^3$.
The scaling factor $(1+t^2)/(1-t^2)$ in (\ref{eq:entjump})
then leads to a constant jump in entropy per vortex per superconducting
layer. In
Ref.~\onlinecite{us} this was found to be $\Delta S_d\approx 0.4k_B$ when YBCO
parameters were used, consistent with the experimental result measured in
Ref.~\onlinecite{Schilling}.

The main point of this paper is to apply similar considerations to 
 the heat capacity. This can be defined for changes at constant field $H$ or
at constant flux density $B$. In the incompressible limit (which coincides
with $\lambda/a_0\rightarrow\infty$) these will be the
same, and
\begin{equation}
  \label{eq:heatb}
C_H= C_B=T\left( \frac{\partial S}{\partial T}\right)_B=
\left( \frac{\partial E}{\partial T}\right)_B.
\end{equation}
(In the next section we will consider the difference \hbox{$C_H-C_B$} for a
compressible system, and show that the difference between the 
jump in specific heat at constant $B$ and at constant $H$ is negligible
on the YBCO  melting line.)
We again use the results
of Ref.~\onlinecite{NordborgBlatter} (shown here  in Fig.~\ref{fig:1})
but this time to determine the jump in the heat capacity.

Let us first assume that $\varepsilon_0$ is independent of temperature. 
Using (\ref{eq:heatb}), the
 heat capacity at constant density takes the form,
\begin{equation}
  \label{eq:c0}
  {C_B}_0  =\frac{\partial}{\partial \tau}\langle f_L\rangle
=\frac{\langle {f_L}^2\rangle -{\langle f_L\rangle}^2}{\tau^2}.
\end{equation}
Note that this ``bare'' heat capacity is simply a measure of the amplitude of
energy fluctuations, which can be expected to jump at the transition from
solid to liquid.
It turns out that the first form is more accurate to use, once many
temperatures have been sampled in the simulations. The change in slopes in
Fig.~\ref{fig:1} gives the result
\begin{equation}
  \label{eq:delc0}
  \Delta {C_B}_0\approx 0.38 NL_z/a_0.
\end{equation}
This
is interpreted as a jump in heat capacity of 0.38~$k_B$ per
vortex degree of freedom (defined above).
As Schilling {\em et al}.\ point out, this
result is far too small to account for the jump in specific heat observed in
their experiment. However, we have neglected the additional terms arising from
the temperature dependent parameters, which we have seen give dominant
contributions to the
entropy jump close to $T_c$.
For a general temperature dependence  of
the energy scale $\varepsilon_0(T)$, we find that the heat capacity is,
\begin{equation}
  \label{eq:cb}
C_B=-\frac{T^2}{\varepsilon_0}\frac{d^2\varepsilon_0}{dT^2}
\frac{\langle f_L\rangle}{\tau} +\left(
  1-\frac{T}{\varepsilon_0}\frac{d\varepsilon_0}{dT} \right)^2 {C_B}_0.
\end{equation}
Note that we need both results (\ref{eq:delc0}) and (\ref{eq:delfl}) 
to find the jump in $C_B$, although as $T_c$ is approached and 
$\varepsilon_0\rightarrow 0$, the last term in (\ref{eq:cb}) dominates.

In the following, we consider a more complex dependence for the energy 
scale of the 
vortex system $\epsilon_0(T)$
than was used in Ref.~\onlinecite{us} and
include a correction that accounts for the suppression in the superconducting
density as the mean-field $B_{c2}$ line is approached.  
This $B_{c2}$ correction is important for a quantitative fit to London theory
of the melting line of YBCO at fields above 1--2~T. While good agreement 
for the entropy jump was
found in Ref.~\onlinecite{us} between the prediction of the uncorrected London
theory  and the experimental values, we will
see that the specific heat jump determined from (\ref{eq:cb})
is more sensitive to the exact form of the melting line. This is
because the terms arising
from the internal temperature dependence of $\varepsilon_0(T)$ are more
strongly diverging in this case than for the entropy jump in 
Eq.~(\ref{eq:entjump}).

In the limit \hbox{$B\approx B_{c2}$}, the
system is described by Ginzburg-Landau (GL) theory, with an order parameter
$\psi$ restricted to the lowest Landau level.\cite{Abrikosov}
The GL free energy
takes the form $F_{GL}=\int d^3r\left[- \alpha |\psi|^2 +\beta |\psi|^4 
+\gamma |\partial\psi/\partial z|^2\right]$, where
$\alpha(T,B)=\alpha_0(1-b)$ and $b=B/B_{c2}(T)$ 
(see for example
Ref.~\onlinecite{RuggeriThouless}). Notice the dimensional reduction in the
derivative term; there is a diverging GL coherence length in the field
direction only.\cite{LeeShenoy} The only length scale remaining in the two
perpendicular directions is the magnetic length $a_0$.
The order parameter scale is 
$|\psi_0|^2=\alpha/2\beta$ and the condensation energy scale is 
$2\alpha|\psi_0|^2=\alpha^2/\beta$. For this reason, the shear modulus has the
behavior $c_{66}\sim (1-b)^2$ (as calculated by 
Labusch\cite{Labusch}). However, the tilt modulus at short wavelengths
(equivalent to the
superfluid density in the $z$-direction) has a linear
dependence\cite{BrandtGLel}  
$c_{44}\sim (1-b)$. (The correct thermodynamic limit of a constant $c_{44}$ as
$k_z\rightarrow 0$ is not captured by the above LLL free energy, which is
restricted to wavelengths below the penetration depth, 
$k_z>1/\lambda$. The full calculation in Ref.~\onlinecite{BrandtGLel} recovers
this limit within GL theory.)

Our intention is to use an {\em effective} pairwise interaction in
(\ref{eq:London}) that gives the London model at low fields, but with the
correct elastic moduli in the high field limit. We take the scale of interactions to be 
$\tilde{\varepsilon}_0(T)=\varepsilon_0(T)(1-b)^2$. This then accounts for the
suppression of the order parameter around both vortex segments in an
interaction term. To account for the increasing stiffness in the field
direction we scale the {\em effective} anisotropy
$\tilde{\varepsilon}=\varepsilon(1-b)^{-1/2}$. Using these effective
parameters in the London model we find $c_{66}\propto b(1-b)^2$, in agreement
with the results of GL theory at intermediate fields that have 
recently been calculated,\cite{Brandt2}
and $c_{44}\propto (1-b)$.
 A similar extrapolation technique was used
twenty years ago by Brandt.\cite{BrandtLel}
Converting the effective London system (with energy scale
$\tilde{\varepsilon}_0$
and anisotropy $\tilde{\varepsilon}$) to an isotropic one
by scaling all lengths in the field
direction, as in Eq.~(\ref{eq:scaling}), 
gives the overall energy scale factor,
\begin{eqnarray}
  \label{eq:e0}
  \varepsilon_0^{\rm eff}=[\tilde{\varepsilon}_0\cdot \tilde{\varepsilon}^2
\tilde{\varepsilon}_0]^{1/2}&=&\varepsilon\varepsilon_{00}
(1-t^2)(1-b)^{3/2}\\
&=& \varepsilon\varepsilon_{00} (1-t^2-\tilde{b})^{3/2}(1-t^2)^{-1/2},\nonumber
\end{eqnarray}
\begin{figure}[b]
\centerline{\epsfxsize= 7.5cm\epsfbox{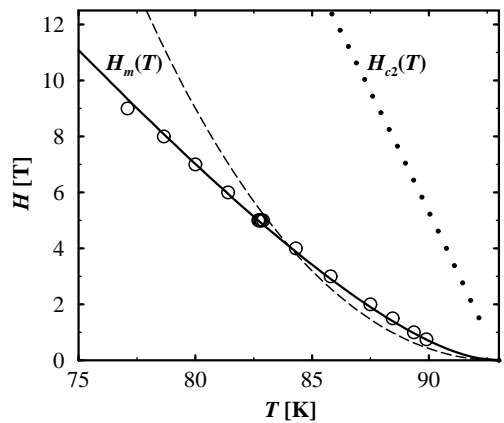}}
\caption{ London scaling fits to the melting line observed in Ref.~3 (circles) 
with (full line) and without (dashed line) $H_{c2}$ corrections.
The $H_{c2}$ line we used in shown as the dotted line.
The parameters used are given in the text.}
\label{fig:2}
\end{figure}
\noindent
with $\varepsilon_{00}=(\Phi_0/4\pi\lambda_0)^2$.
In the last line we have assumed the form $B_{c2}(T)=B_{c2}(0)(1-t^2)$ and
written
$\tilde{b}=B/B_{c2}(0)$.

We now compare our results to the measurements of Schilling 
{\em et al.}\cite{Schilling,Schilling2} 
and concentrate on the case where the field is
directed perpendicular to the CuO$_2$ layers. 
 In the scaling regime, the melting line is described by a
fixed value of $\tau=\tau_m$, i.e. 
$T_m=g a_0(B)\varepsilon_0^{\rm eff}(T_m,B)$. In Fig.~\ref{fig:2} 
we show our fit to the experimental melting line, where we find a value of 
$g\approx 0.21$ (this corresponds to a Lindemann number of $c_L=0.24$
when the melting line from the Lindemann criterion is written 
in the standard form\cite{review} of $T_m\approx 2\sqrt{\pi}c_L^2
\varepsilon\varepsilon_0 a_0$).
To obtain this fit we took the physical values: 
$T_c=93~\hbox{K}$,
$\varepsilon=1/8$,
$dH_{c2}/dT |_{T=T_c}=1.8\hbox{~TK}^{-1}$,
$\lambda_0=1300\hbox{~\AA}$. In Fig.~\ref{fig:2} we also show our fit without
including $B_{c2}$ corrections, which leads to a value of $g=0.13$, or
$c_L=0.19$.

\begin{figure}[b]
\centerline{\epsfxsize= 8.5cm\epsfbox{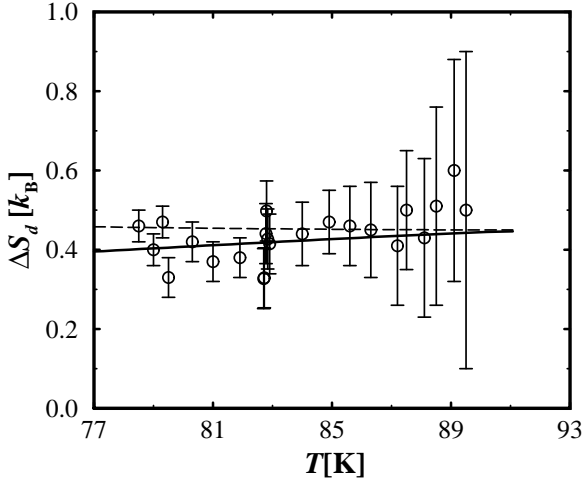}}
\caption{The calculated entropy jump per vortex per layer (solid line). For
  comparison the results when $B_{c2}$ corrections are not included are also
  shown (dashed line). Both curves are within the error bars of the
  experimental values from Ref.~3.  }
\label{fig:3}
\end{figure}

With (\ref{eq:e0}) the formula for the entropy jump in (\ref{eq:entjump}) now
becomes,
\begin{equation}
  \label{eq:entjumpbc2}
  \Delta S
=\frac{[1-\tilde{b}+(2\tilde{b}-t^2)t^2]}{(1-t^2-\tilde{b})(1-t^2)}\Delta S_0.
\end{equation}
Note that the factor on the RHS now diverges at the $B_{c2}$ line rather than
at $T_c$.
In Fig.~\ref{fig:3} we compare our predictions for the entropy jump per vortex
per layer,
calculated with and without $B_{c2}$ corrections, to the experimental values
of Ref.~\onlinecite{Schilling}. Interestingly, there is little difference
between the two approaches in this temperature regime.

Substituting (\ref{eq:e0}) into (\ref{eq:cb})
leads to
\begin{eqnarray}  \label{eq:cbcorr}
 \Delta C_B&&=
\frac{[1-\tilde{b}+t^2(2\tilde{b}-t^2)]^2}{[(1-t^2)(1-t^2-\tilde{b})]^2}\, 
\Delta {C_B}_0 +\\
&& +\, 
\frac{t^2[2(1-t^2)^3-\tilde{b}(1-t^2)^2-\tilde{b}^2(1+2t^2)]}{ 
[(1-t^2) (1-t^2-\tilde{b})]^2}\,\,
\Delta \frac{\langle f_L\rangle}{\tau}.\nonumber
\end{eqnarray}
The result for $\Delta C_B$ in (\ref{eq:cbcorr}) combined with (\ref{eq:delc0})
and (\ref{eq:delfl}) tells us the jump in the heat capacity
{\em per degree of freedom} (i.e., per volume $V_{\rm edf}$), which
 diverges as $T_c$ is approached. Indeed, we find that 
over the region of temperatures where measurements have
been performed, the heat capacity jump is of order $100\, k_B$ 
per vortex degree
of freedom. Because the melting field falls to zero, the volume of
a degree of freedom also diverges in this limit, such that
the heat capacity jump in
a physical sample of fixed volume drops to zero, $\Delta C_B\propto(1-t^2)$,
on approaching $T_c$.
 In Ref.~\onlinecite{Schilling2} the results for the jump in heat capacity 
are expressed for a fixed volume. 
The 
{\em specific} heat is defined as the heat capacity in a
mole of YBCO which fills a volume 
 $V_{\rm mol}=1.05\times 10^{-4}\hbox{m}^3$
(this is the volume of $6.02\times 10^{23}$ unit cells). We therefore 
convert to the specific heat jump $\Delta c_B =(V_{\rm mol}/V)\Delta C_B$
and make use of the fit to the melting line of Fig.~\ref{fig:2}
in Eq.~(\ref{eq:delc0}),
$\Delta C_{B0}\approx 0.38 V/V_{\rm edf}\propto B_m^{3/2}(T)$.

\begin{figure}[b]
\centerline{\epsfxsize= 8.5cm\epsfbox{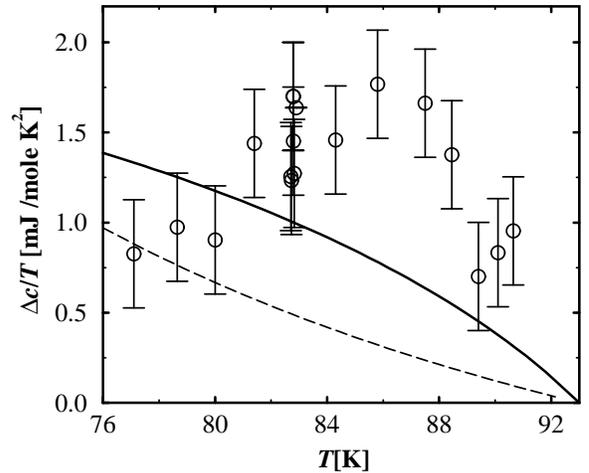}}
\caption{ The calculated jump in specific heat (solid line) with the
  units used in Ref.~4, and the measured points from that paper.
In this case there is a significant difference from the calculation ignoring
$B_{c2}$ corrections (dashed line)}
\label{fig:4}
\end{figure}

Our result for the specific heat jump per molar volume is shown in
Fig.~\ref{fig:4} along with the experimentally measured points.
This shows that we can explain the size of the specific heat jump within about
50\% accuracy. We should not expect a much better agreement
as we 
extrapolate the London approximation into a region where it will not be
exact and also neglect the effects of disorder. 
In fact, the measurements of the
magnetization jump in Ref.~\onlinecite{Welp}  show $\Delta B$ falling to zero
at a temperature several Kelvin below $T_c$, and this is usually attributed to
the effects of quenched disorder which may smooth the jumps at a first-order
transition.\cite{Geshkenbein} There may be a similar explanation behind the
non-monotonic behavior in the experimental value of $\Delta c$.
Nevertheless, the important point here is that the size of the
experimentally observed jumps
in the 
specific heat are consistent with a model that only includes field induced
vortices. 

\section{The magnetization and its derivatives}\label{sec:mag}

The aim of this section is to use our results for the entropy and specific
heat jumps from the previous section to calculate the jumps in the derivatives
of $B$ (or in the magnetization $M=(B-H)/4\pi$) with respect to $H$ and $T$.
We first introduce the Clausius-Clapeyron relation, and then describe the
conditions under which the jump in entropy in a system at constant external
field $H$ is the same as the entropy jump in a system of fixed flux density
$B$. We need these conditions to hold in order to justify our application of
the results of Section~\ref{sec:heat}, which we calculated at constant $B$, to
the experimental results, which are measured at constant $H$.
After ensuring that these conditions are fulfilled for the melting line of
YBCO, we then calculate the jumps in the thermal compressibility 
$\Delta (\partial B/\partial T)_H$ and in the susceptibility
$\Delta (\partial B/\partial H)_T$. Finally we will be in a position to
describe the conditions for the specific heat jump to be approximately the
same in a constant $B$ system compared to that in a constant $H$ system, and
then verify that these conditions are also satisfied.  

\begin{figure}[b]
\centerline{\epsfxsize=8.5cm\epsfbox{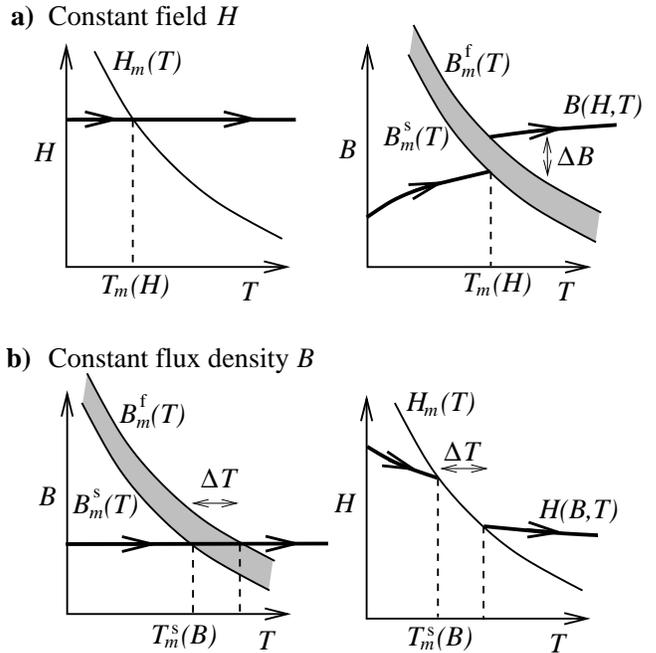}}
\vspace{0.6cm}
\caption{ Schematic phase diagrams in the $H$-$T$ plane, showing the phase 
boundary $H_m(T)$, and the $B$-$T$ plane, showing a region of coexistence of 
the fluid and lattice phases between
the two boundaries $B^{\rm s}(T)$ and $B^{\rm f}(T)$.
a) A path is shown for increasing temperature at constant external $H$,
crossing a sharp phase transition at $T_m(H)$. In the $B$-$T$ plane, the
coexistence region is crossed at constant $T$, with a jump in induction
$\Delta B|_H$. b) The path for increasing temperature at constant $B$ is shown.
While traversing the coexistence region, the path remains on the phase
boundary $H_m(T)$ in the $H$-$T$ plane. There is a well defined width of the
transition, $\Delta T|_B$. 
}
\label{fig:5}
\end{figure}

\begin{figure}[b]
\centerline{\epsfxsize= 6cm\epsfbox{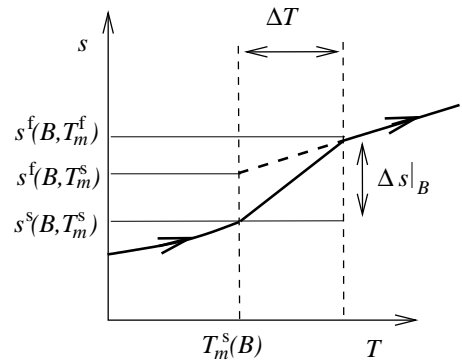}}
\vspace{0.6cm}
\caption{ Schematic graph of the entropy as the melting transition is crossed
  at constant flux density $B$. The entropy jump is defined across the whole
  width of the transition, 
$\Delta s|_B=s^{\rm f}(B,T_m^{\rm f})-s^{\rm s}(B,T_m^{\rm s})$.}
\label{fig:6}
\end{figure}

At the first-order transition of a system at constant field $H$, the continuity
of the Gibbs free energy $G=F-(BH/4\pi)V$ leads to the Clausius-Clapeyron
equation
\begin{equation}
  \label{eq:cc}
  \Delta s=-\frac{1}{4\pi}\frac{dH_m}{dT}\Delta B,
\end{equation}
relating the jump in flux density $\Delta B$ to the melting line $H_m(T)$
and the jump in the entropy density
$\Delta s$. 
In the previous section we calculated  the jumps in entropy and
heat capacity for a transition at fixed $B$, using the scaling form of the
London model [see Eq.~(\ref{eq:entjump})]. An important difference between the
melting transition at constant $B$ and at constant $H$ is that in the former,
coexistence occurs over a finite temperature region, see Fig.~\ref{fig:5}.
The entropy jumps at constant $H$ and constant $B$  will not be the same:
We must take into account the
different values of $B$ on either side of the constant $H$ 
transition,
$\Delta B|_H=B^{\rm f}(H,T_m)-B^{\rm s}(H,T_m)$, and also the width in
temperature of the constant $B$ transition, 
$\Delta T|_B=T_m^{\rm f}(B)-T_m^{\rm s}(B)$
 [the symbol s denotes the
solid (lattice) phase, while the symbol f is for the fluid (liquid) phase]. 
The entropy jump at constant $B$  is
defined as $s^{\rm f}(B,T_m^{\rm f})-s^{\rm s}(B,T_m^{\rm s})$ as shown in
Fig.~\ref{fig:6}, which to first order in $\Delta T$ is
\begin{equation}
  \label{eq:entdef}
\Delta s|_B=
s^{\rm f}(B,T_m^{\rm s})-s^{\rm s}(B,T_m^{\rm s})
+\frac{c_B^{\rm f}}{T_m^{\rm f}}\Delta T|_B.
\end{equation}
\noindent
We can now relate the entropy jump at fixed $H$ to that at fixed $B$,
\begin{eqnarray}
  \label{eq:entbenth}
  \Delta s|_H&=&s^{\rm f}(H_m,T)-s^{\rm s}(H_m,T)\nonumber\\
&=& s^{\rm f}(B_m^{\rm s},T)-s^{\rm s}(B_m^{\rm s},T)+
\left( \frac{\partial s^{\rm f}}{\partial B} \right)_T\Delta B|_H\nonumber\\
&=&\Delta s|_{B_m^{\rm s}} +
\frac{1}{4\pi}\left( \frac{\partial H^{\rm f}}{\partial B} \right)_T
\left( \frac{\partial B^{\rm f}}{\partial T} \right)_H \Delta B|_H
-\nonumber\\
&&-\frac{c_B^{\rm f}}{T}\Delta T|_{B_m^{\rm s}},
\end{eqnarray}
where in the last line we have used the Maxwell relation
\begin{equation}\label{eq:max}
\left(\frac{\partial s}{\partial B}\right)_T=-\frac{1}{4\pi}
\left(\frac{\partial H}{\partial T}\right)_B=\frac{1}{4\pi}
\left(\frac{\partial H}{\partial B}\right)_T
\left(\frac{\partial B}{\partial T}\right)_H.
\end{equation}
 Equation (\ref{eq:entbenth}) shows that there are
two terms that lead to corrections in the
 entropy jump at fixed $B$ compared to fixed
$H$. The physical explanation is that the latent heat at fixed $H$ includes
the magnetic
work done to increase the flux density by $\Delta B|_H$, while at fixed $B$
there is an extra heat increase due to the temperature increase across the
transition $\Delta T|_B$. Our results  for the
entropy jump at constant $B$ will only be valid at constant $H$, 
\hbox{$\Delta s|_H\approx\Delta s|_{B=B(H)}$}, if the two conditions 
\begin{eqnarray}
  \label{eq:conddelt}
\frac{c_B^{\rm f}}{T}\Delta T|_{B_m^{\rm s}}&\ll& \Delta s|_H\\
  \label{eq:conddelB}
\frac{1}{4\pi}
\left( \frac{\partial H^{\rm f}}{\partial B} \right)_T
\left( \frac{\partial B^{\rm f}}{\partial T} \right)_H \Delta B|_H
  &\ll&\Delta s|_H
\end{eqnarray}
are fulfilled.

To apply the results of the previous section, we must check the above
conditions for the case of vortex-lattice melting in YBCO.
We first consider the term arising from the temperature increase across the
constant $B$ melting transition.
By writing
$\Delta B|_H\approx (dB_m/dT) \Delta T|_B$ (see Fig.~\ref{fig:5}), 
and using the Clausius-Clapeyron
relation (\ref{eq:cc}),
we can express the condition (\ref{eq:conddelt}) in the form, 
\begin{equation}
  \label{eq:conddelt2}
 \frac{c_B^{\rm f}}{T} \ll \frac{1}{4\pi}\left(\frac{dB_m}{dT}\right)^2. 
\end{equation}
Using the numerical results presented in
 the previous section, the total heat capacity at
the YBCO melting transition is of order $c_B/T\approx 10$~Jm$^{-3}$K$^{-2}$.
From Fig.~\ref{fig:2} we have $dH_m/dT\approx -0.7$TK$^{-1}$ on the melting
line of YBCO, which gives 
$(dH_m/dT)^2/4\pi\approx 4\times 10^5$~Jm$^{-3}$K$^{-2}$. 
Therefore this condition is well satisfied
on the YBCO melting line. Finally, we consider the term involving the jump
in flux density at the transition with fixed $H$.
If we insert the Clausius-Clapeyron relation (\ref{eq:cc}) into the RHS of
condition (\ref{eq:conddelB}) we find that the
condition is equivalent to $(\partial B^{\rm f}/\partial T)_H\ll dH_m/dT$,
which is to say that the thermal compressibility of the vortex system 
$(\partial B^{\rm f}/\partial T)_H = 4\pi(\partial M^{\rm f}/\partial T)_H$
must be
smaller than the slope of the melting line. This
is satisfied as long as the  magnetization (which is of order the lower
critical field $H_{c1}$) is less than the melting field, $H_{c1}\ll H_m$, which
is the case for vortex lattice melting in YBCO.
We have therefore confirmed in our regime of interest (the large field
scaling limit) that the same entropy jump is found for a transition at 
constant $H$ and at constant $B$.
If we also take $B_m(T)=H_m(T)$, we can calculate $\Delta B$  
in the scaling regime using the Clausius-Clapeyron relation (\ref{eq:cc}). 
This reasoning was used in
Ref.~\onlinecite{us} and excellent agreement was found between the London
model predictions and the measured values of $\Delta B$ in YBCO by 
Welp {\em et al.}\cite{Welp,Schilling}

We now calculate the jumps in the derivatives of the magnetization.
Consider a quantity $X(T,H)$ which is discontinuous at the melting field
$H_m(T)$ with a jump $\Delta X(T)$. The total temperature derivative of the
jump in $X$ (defined along the melting line) is then
\begin{equation}
  \label{eq:dxdt}
 \frac{d\Delta X}{dT}= \Delta\left( \frac{\partial X}{\partial T}\right)_H 
+\frac{dH_m}{dT} \Delta\left( \frac{\partial X}{\partial H}\right)_T.
\end{equation}
Note that inserting $X=G$, the Gibbs free energy which is continuous,
simply gives the Clausius-Clapeyron equation.
 Applying (\ref{eq:dxdt}) to the entropy jump by
taking $X=s$, and using
the Maxwell relation $4\pi(\partial s/\partial H)_T
=(\partial B/\partial T)_H$, we find
\begin{equation}
  \label{eq:cc2}
  \frac{1}{4\pi}\frac{dH_m}{dT}
\Delta\left( \frac{\partial B}{\partial T}\right)_H
=-\frac{\Delta c_H}{T} +\frac{d\Delta s}{dT}.
\end{equation}
This allows us to find $\Delta(\partial B/\partial T)_H$ once we know the
corresponding discontinuities in the calorimetric quantities $c_H$ and $s$.
In the SQUID measurements of the magnetization in Ref.~\onlinecite{Welp},
 an experimental value is found at a field of $H=4.2$T of 
$\Delta(\partial B/\partial T)_H=
4\pi\Delta(\partial M/\partial T)_H\approx 0.2$~GK$^{-1}$ 
(the accuracy is only of order 50\%).
We convert from magnetic energy units to calorimetric units using
$1\,\hbox{erg cm}^{-3}\equiv 0.1\,\hbox{J m}^{-3}=0.011\,\hbox{mJ mol}^{-1}$ 
for YBCO.
For the specific heat jump we take
 $\Delta c_H\approx\Delta c_B= 1\hbox{ mJ mol}^{-1}{\rm K}^{-2}$ 
from Fig.~\ref{fig:4}. We calculate 
$\Delta s=\Delta S/V$ from Eqs.~(\ref{eq:entjump}) and (\ref{eq:delfl})
giving $d\Delta s/dT=-0.35\,\hbox{mJ mol}^{-1}{\rm K}^{-2}$ at a field of
$H_m=4.2$T. Inserting these results into (\ref{eq:cc2}) along
with the melting slope of $dH_m/dT=-0.7\,$TK$^{-1}$, 
we find $\Delta(\partial B/\partial T)_H=0.23\,$GK$^{-1}$ consistent with the
experimental value at this field. 

We now use two further relations between these jumps which will allow the jump
in the susceptibility, $\Delta(\partial B/\partial H)_T$, to be calculated.
The total derivative with respect to temperature 
of (\ref{eq:cc}) may be written as
\begin{eqnarray}
  \label{eq:cc3}
  \Delta c_H=-\frac{T}{4\pi}&&\left[2\frac{dH_m}{dT}\Delta 
\left(\frac{\partial B}{\partial T}\right)_H 
+\right. \nonumber\\
&&\left. +\left(\frac{dH_m}{dT}\right)^2
\Delta\left(\frac{\partial B }{\partial H}\right)_T
 +\frac{d^2H_m}{dT^2}\Delta B\right].
\end{eqnarray}
This equation was used in Ref.~\onlinecite{Schilling2} to successfully test the
thermodynamic consistency of the observed specific heat jump with the
magnetization measurements of Welp {\em et al.}\cite{Welp}
Combining this with (\ref{eq:cc2}) gives an equation for the jump in 
$(\partial B/\partial H)_T$ in terms of calorimetric quantities only,
\begin{eqnarray}
  \label{eq:cc4}
\frac{1}{4\pi}  \left(\frac{dH_m}{dT}\right)^3
\Delta\left(\frac{\partial B }{\partial H}\right)_T=&&
\frac{dH_m}{dT} \left(\frac{\Delta c_H}{T} -2\frac{d\Delta s}{dT}\right)+
\nonumber\\
&&+\frac{d^2H_m}{dT^2}\Delta s.
\end{eqnarray}
Note that this may also be derived by inserting $X=B$ in Eq.~(\ref{eq:dxdt})
and using the Clausius-Clapeyron relation.
We find this jump along the melting line of YBCO to be of order 
$\Delta(\partial B/\partial H)_T\approx 10^{-5}$G/Oe, 
consistent with the assumption
that $B\approx H$. However, we must be careful about ignoring this
term. For instance, in the scaling regime, and from experimental values in
YBCO, 
the temperature derivative $d(\Delta B)/dT=\Delta(\partial B/\partial T)_H
+(dH_m/dT)\Delta(\partial B/\partial H)_T$ is negative but the jump
$\Delta(\partial B/\partial T)_H$ 
is positive and therefore smaller in magnitude
than $(dH_m/dT)\Delta(\partial B/\partial H)_T$.

We now determine whether the jump in the heat capacity at constant $H$, as
measured in experiments on YBCO, may be approximated by the jump in heat
capacity at constant $B$, which was calculated in
Section~\ref{sec:heat}. As in our earlier consideration of the entropy jump,
we must include the change in $B$ at a constant $H$ transition as well as the
finite temperature width of a constant $B$ transition. However, there is an
extra difference in that, while the entropy is a function of state,
$S(H,T)=S[B(H,T),T]$, the heat capacity is defined differently for changes at
constant $B$ or at constant $H$, 
\begin{eqnarray}
 \label{eq:heath}
c_H=T\left( \frac{\partial s}{\partial T}\right)_H&=&
T\left( \frac{\partial s}{\partial T}\right)_B+
T\left( \frac{\partial B}{\partial T}\right)_H
\left( \frac{\partial s}{\partial B}\right)_T\nonumber\\
&=&
c_B+\frac{T}{4\pi}\left(\frac{\partial H}{\partial B}\right)_T
\left(
\frac{\partial B}{\partial T}\right)_H^2,
\end{eqnarray}
where the last line again uses the Maxwell relation (\ref{eq:max}).
With the same reasoning that lead to (\ref{eq:entbenth}) and using
the notation defined as Fig.~\ref{fig:5} we have,
\begin{equation}
  \label{eq:delcbh}
  \Delta c_H|_B=\Delta c_H|_H 
-\left( \frac{\partial c^{\rm f}_H}{\partial B}\right)_T
\Delta B|_H
+\left( \frac{\partial c^{\rm f}_H}{\partial T}\right)_B
\Delta T|_B.
\end{equation}
We can also find the jump in the difference $(c_H-c_B)$ from (\ref{eq:heath})
by expanding to first order in the jumps,
\begin{eqnarray}
  \label{eq:delchcb}
  \Delta (c_H-c_B)|_B\approx 
&&\frac{T}{4\pi}  \left[ -
\left(\frac{\partial B}{\partial T}\right)_H^2 
\Delta \left(\frac{\partial B}{\partial H}\right)_T +\right.
\\&& +\left.2 
\left(\frac{\partial H}{\partial B}\right)_T
\left(\frac{\partial B}{\partial T}\right)_H
\Delta \left(\frac{\partial B}{\partial T}\right)_H \right].\nonumber
\end{eqnarray}
Our previous results show that the first term on the RHS is negative, while
the second term is positive. Although the total $c_H-c_B$ is rigorously
positive for thermodynamic stability, the change in this quantity may be of
either sign.
Combining this equation (\ref{eq:delchcb}) 
with (\ref{eq:delcbh}) gives
\begin{eqnarray}
  \label{eq:delcbdelch}
  \Delta c_B|_B&=&\Delta c_H|_H 
+\left( \frac{\partial c^{\rm f}_H}{\partial T}\right)_B
\Delta T|_B
 -\left( \frac{\partial c^{\rm f}_H}{\partial B}\right)_T
\Delta B|_H
 +\nonumber\\
&&+\frac{T}{4\pi}
\left(\frac{\partial B}{\partial T}\right)_H^2
\Delta\left(\frac{\partial B}{\partial H}\right)_T
-\nonumber\\
&&-\frac{2T}{4\pi} 
\left(\frac{\partial H}{\partial B}\right)_T
\left(\frac{\partial B}{\partial T}\right)_H
\Delta\left(\frac{\partial B}{\partial T}\right)_H.
\end{eqnarray}
We can group the corrections to $\Delta c_B|_B$ into two contributions. The
first  correction term in (\ref{eq:delcbdelch}) can be ignored compared to the
total specific heat jump if $(\partial c_H^{\rm f}/\partial T)\Delta T|_B
\ll\Delta c_B$.
After some manipulations with the
Clausius Clapeyron relation it may be written as
\begin{equation}
  \label{eq:conddelcdelt}
 \left(  \frac{\partial c_B^{\rm f}}{\partial T}\right)_B \Delta s
\ll \frac{1}{4\pi}\left(\frac{dB_m}{dT}\right)^2 \Delta c_B,
\end{equation}
which we find to be satisfied in the scaling regime from the simulation
results. A second condition is found from the three remaining correction terms
in (\ref{eq:delcbdelch}) after dropping the term  $\propto\Delta T|_B$. 
Setting 
$(\partial H/\partial B)_T=1$  and using the identity 
$(\partial c_H/\partial B)_T=-T(\partial H/\partial B)_T
(\partial^2 B/\partial T^2)_H$ [which also arises from the Maxwell relation
(\ref{eq:max})], gives
\begin{eqnarray}
  \label{eq:again}
\Delta c_H|_H  - \Delta c_B|_B
&&=\frac{-T}{4\pi}\left[
-2 \left(\frac{\partial B}{\partial T}\right)_H \Delta 
\left(\frac{\partial B}{\partial T}\right)_H+\right.\\
&& \hspace{-1cm}\left. + \left(\frac{\partial B}{\partial T}\right)_H^2\Delta 
\left(\frac{\partial B}{\partial H}\right)_T
 +\left(\frac{\partial^2B}{\partial T^2}\right)_H
\Delta B
\right].\nonumber
\end{eqnarray}
A direct comparison of this equation with the differentiated
Clausius-Clapeyron relation (\ref{eq:cc3})
shows that we can take $\Delta c_H|_H\approx\Delta c_B|_B$ as long as
$|dH_m/dT|\gg (\partial B/\partial T)_H$, which is the same condition we found
for comparing the entropy jump at constant $H$ versus constant $B$, and which
is valid as long as $H_m\gg H_{c1}$.
Therefore, our calculation of
$\Delta c_B$ at constant $B$ is justified in the application to vortex lattice
melting in YBCO.

\section{Different field angles}\label{sec:angle}

In Ref.~\onlinecite{Schilling2} the heat capacity in YBCO
was measured both
with the magnetic field  parallel
 to the c-axis, $\hbox{\bf H}\parallel\hbox{\bf c}$, and 
parallel to the superconducting layers, 
$\hbox{\bf H}\perp\hbox{\bf c}$. 
More recent experiments have been performed for a range of angles between
these two limits.\cite{Schilling3}
In the scaling regime, the effects of
rotating the field may be simply understood using the anisotropic scaling
rules of Blatter, Geshkenbein, and Larkin.\cite{BlattGeshLark} The basis of
these scaling rules is that the anisotropic system can be transformed to an
isotropic one by scaling the lengths along the {\bf c}-axis. 
For a magnetic field enclosing an angle $\theta$ with the {\bf c}-axis, 
a given
quantity $Q$  then follows the scaling law
\begin{equation}
  \label{eq:anis}
  Q(\theta,H,T,\lambda_{ab},\varepsilon)=s_Q \tilde{Q}(\varepsilon_\theta H, 
T/\varepsilon,\lambda),
\end{equation}
where $\tilde{Q}$ is the corresponding quantity in the isotropic
superconductor (with $\lambda=\lambda_{ab}$), and 
$\varepsilon_\theta^2 =\varepsilon^2\sin^2\theta+\cos^2\theta$. 
The factor $s_Q$
depends on the quantity $Q$. For the scalar quantities
volume and energy it is independent of the angle, $s_Q=\varepsilon$, but for
magnetic fields it is equal to $s_B=1/\varepsilon_\theta$. 

An initial consequence of (\ref{eq:anis}) is that, at a fixed temperature, the
magnetic field at which the system has the same physics scales as 
$H_X(\theta)=\tilde{H}_X/\varepsilon_\theta$. This implies that
 the melting line
is raised to higher fields as the angle $\theta$ is increased,
\begin{equation}
\label{eq:hmscal}
H_m(T,\theta)=H_m(T,0)/\varepsilon_\theta.
\end{equation}
 The same scaling form applies to
 the upper critical field $H_{c2}(T,\theta)$. These fields are lower
for $\hbox{\bf H}\parallel\hbox{\bf c}$
 than for $\hbox{\bf H}\perp\hbox{\bf c}$ by a
factor $\varepsilon$. 
Furthermore, if we take the quantity $Q$ in (\ref{eq:anis}) to be the energy,
entropy or heat capacity of a fixed volume (e.g.\ the volume of the crystal)
then after scaling the magnetic field as in (\ref{eq:hmscal}) this quantity
will be {\em independent} of the angle $\theta$. Therefore, if we write the
entropy jump as a function of melting temperature, $\Delta s(T,\theta)$,
the following relation holds, 
\begin{equation}
  \label{eq:delsofttheta}
  \Delta s[T_m(H,\theta),\theta]=\Delta s[T_m(\varepsilon_\theta H,0),
0].
\end{equation}
We can also replace $s$ by $c$ in this equation.
Therefore the measured jumps
in the specific heat and in the entropy per molar volume, when plotted against
the melting temperature, will be identical for the different angles. This
prediction is fulfilled to within experimental error in recent measurements by
Schilling.\cite{Schilling3}

To understand this result in more detail, we now consider the results
for the jumps 
(\ref{eq:entjump}) and
(\ref{eq:cbcorr}). Notice that both of these results are of the form 
$\Delta X = f(T/T_c,B/B_{c2}) V/V_{\rm edf}$. The scaling rule (\ref{eq:anis}) 
tells us that at fixed $T$, the reduced variables $t=T/T_c$
and $b=B_m/B_{c2}$ at melting are independent of $\theta$. We therefore
concentrate on the angular dependence of $V_{\rm edf}$. In the isotropic
system, 
$\tilde{V}_{\rm edf}=\tilde{a}_0^3$. However, the three perpendicular
dimensions of $V_{\rm edf}$ will scale differently with $\theta$.
If the field lies in the $y$-$z$ plane
then 
$l^{\rm edf}_x=\tilde{l}=\tilde{a}_0$, 
whereas the lengths in the plane of rotation scale 
as,\cite{review}
 \begin{eqnarray}
   \label{eq:lensc}
   l^{\rm edf}_l&=&\frac{\varepsilon}{\varepsilon_\theta}\tilde{l},\,\,\,\,\,
\hbox{\bf l}_l\parallel \hbox{\bf B},
\nonumber\\
   l^{\rm edf}_t&=&\varepsilon_\theta\tilde{l}, \,\,\,\,\,
\hbox{\bf l}_t\perp \hbox{\bf B}.
 \end{eqnarray}
Therefore, at fixed temperature, but scaling the magnetic field as in
Eq.~(\ref{eq:hmscal}), we find that the volume per degree of freedom stays
constant,
\begin{equation}
  \label{eq:vedfscal}
  V_{\rm edf}(T_m,\theta)=l_x^{\rm edf}l_t^{\rm edf}l_l^{\rm edf}
  =\varepsilon {\tilde{l}}^3=\varepsilon {\tilde{a}_0}^3
=\varepsilon {a_0}^3|_{\theta=0}.
\end{equation}
So we see that it is because the volume per vortex degree of freedom is
independent of $\theta$ at the melting field of a fixed temperature
that $\Delta s(T)$ and $\Delta c(T)$ are also independent of $\theta$.

We now consider the case of fixing the magnetic field, and comparing the jumps at
different melting temperatures as the angle of the field is rotated. In this
case, the field scales to a {\em different} value,
$\tilde{H}=\varepsilon_\theta H$,
in the isotropic system for
different values of $\theta$. The melting temperature is,
\begin{equation}
  \label{eq:tmscal}
  T_m(H,\theta)=\varepsilon\tilde{T}_m(\varepsilon_\theta H).
\end{equation}
This equation differs from (\ref{eq:hmscal}) in that
we need to know the full melting
curve in the isotropic case to find the angular dependence of the melting
temperature. Similarly, to determine the angular dependence of the entropy
jump as a function of $H$ from (\ref{eq:delsofttheta}), we need to know the
full function $\Delta s(T,\theta=0)$ and the melting curve $T_m(H,\theta=0)$.
However, in the case of no $B_{c2}$ 
corrections we have an analytic form for the
melting curve, $B_m(T,\theta)\propto (1-t^2)^2/\varepsilon_\theta$, 
and in this limit we can find the
angular scaling at the same values of $H$.
When we stay on the melting line at the same $H_m$, the volume 
$V_{\rm edf}(\theta)=\varepsilon\tilde{V}_{\rm edf}$ scales as
\begin{equation}
  \label{eq:vedfscalh}
  V_{\rm edf}(H_m,\theta)=\varepsilon_\theta^{-3/2}V_{\rm edf}(H_m,0)
\approx \varepsilon\left(\frac{\Phi_0}{B_m\varepsilon_\theta}\right)^{3/2}.
\end{equation}
If the jump in the  entropy and the heat capacity were constant per vortex
degree of freedom then this equation implies that the jumps would scale 
proportional to $\varepsilon_\theta^{3/2}$, as is stated in
Ref.~\onlinecite{Schilling2}.  However, we showed in Section~\ref{sec:heat}
that the temperature dependence of $\varepsilon_0(T)$ gives a non-trivial
temperature dependence to the jumps per vortex degree of freedom.
In the limiting case of no $B_{c2}$ corrections, 
the jump in entropy density is 
$\Delta s\propto (1/V_{\rm edf})(1-t^2)^{-1}\propto B_m\varepsilon_\theta$ 
and the jump in heat capacity is $\Delta c\propto 
(1/V_{\rm edf})(1-t^2)^{-2}\propto (B_m\varepsilon_\theta)^{1/2}$.
This leads to the angular scaling
\begin{eqnarray}
  \label{eq:delsdelcsc}
  \Delta s(H,\theta)= \Delta s(H,0)\varepsilon_\theta\nonumber\\
  \Delta c(H,\theta)= \Delta c(H,0)\varepsilon_\theta^{1/2}.
\end{eqnarray}
For the general
case when $B_{c2}$ corrections are included, there is no such simple form.
However, the more general angular scaling of the jumps (\ref{eq:delsofttheta})
still holds and it is straightforward to find 
$\Delta s(H,\theta)$ and $\Delta c(H,\theta)$ numerically.

\section{Conclusions}

We have not in this paper considered the case of the strongly layered
high-$T_c$ superconductors such as BiSCCO, where the low melting field
$B_m<\Phi_0/\lambda^2$ means that the scaling form of (\ref{eq:scaling}) 
does not apply. Still,
in Ref.~\onlinecite{us},  we successfully determined
the magnetization jump in BiSCCO using dimensional estimates, together with a
Lindemann analysis of the melting line. The reason this works, while simple
estimates for the entropy jump do not, is that the induction is a derivative of
the Gibbs free energy with respect to $H$ (rather than $T$ as for the entropy)
and the $H$ dependence of the vortex parameters is much weaker than the $T$
dependence. The entropy jump for BiSCCO could then 
be found by combining our
result for $\Delta B$ with the melting line and the Clausius--Clapeyron
relation (\ref{eq:cc}). It is not possible to simply follow this procedure to
find $\Delta c$ using the relation (\ref{eq:cc3}), as the simple estimates
will be incorrect for $\Delta(\partial B/\partial T)_H$.
It is also interesting to consider whether
the conditions for the calorimetric jumps to be the same in a constant $B$ 
system and in a constant $H$ system, derived in Sec.~\ref{sec:mag}, hold at
these low fields. The melting field in BiSCCO is of the order of
the bulk lower critical field $H_{c1}$, which is also the  size 
of the magnetization: $H_m\sim H_{c1}\sim 4\pi M$. Therefore the slopes
$dH_m/dT$ and $(\partial B/\partial T)_H$ will be of a similar size 
so that the condition (\ref{eq:conddelB}) 
is not fulfilled 
and the RHS of (\ref{eq:again}) is not small.
In experimental measurements on BiSCCO (at constant $H$) there will
be a significant contribution to the entropy and specific heat jumps 
that is not present in a system of fixed flux density $B$.

To summarize, 
we have calculated the specific heat jump at the melting transition of the
vortex lattice for YBCO, using the scaling form of the London model at high
fields and numerical results from recent simulations. 
We find a jump with the same magnitude as the measured experimental
values. We have used analogous relations to the Clausius--Clapeyron equation
to find the jumps in the magnetization slopes, and these are also
consistent with measured values for YBCO.
A careful analysis was made to make sure that the jumps in the entropy and in
the specific heat are the same for a system at constant $B$ and at constant
$H$. We have constructed the relevant conditions and have verified that they
are satisfied at the vortex--lattice melting line in YBCO.
We have analyzed the changes to the jumps when the 
field angle is rotated with respect to the crystal and thereby
explain the recently
measured angular scaling of the entropy and specific heat jumps in
YBCO.\cite{Schilling3} 
In conclusion, we find that the London model with temperature and field
dependent parameters gives a consistent picture of
the first--order transition observed in YBCO.\cite{MooreChin}

We thank A. Koshelev,
M. Moore, and A. Schilling, for stimulating
discussions, and the Swiss 
National Foundation for financial support.

\end{multicols}

\end{document}